\documentclass[aps,pre,preprint,groupedaddress,showpacs]{revtex4-1}
\usepackage{amssymb}

\usepackage[dvips]{graphicx}
\usepackage{textcomp}
\usepackage{amssymb}

\begin{document}

\title{
Speeding up parallel tempering simulations}

\author{Martin Hasenbusch}
\email[]{Martin.Hasenbusch@physik.hu-berlin.de}
\author{Stefan Schaefer}
\email[]{Stefan.Schaefer@physik.hu-berlin.de}
\affiliation{Institut f\"ur Physik, Humboldt-Universit\"at zu Berlin,
Newtonstr. 15, 12489 Berlin, Germany}

\date{\today}

\begin{abstract}
We discuss  methods that allow to increase the step-size in a parallel
tempering simulation of statistical models and test them  at
the example of the three-dimensional  Heisenberg spin glass. We find 
an overall speed-up of about two for contemporary lattices.
\end{abstract}

\maketitle

\section{Introduction}

The Monte Carlo simulation of statistical models with a rugged free energy
landscape is notoriously difficult. At low temperatures the simulation might
get stuck in one of the valleys of the free energy, which leads to very
large auto-correlation times. One way to overcome this problem is the so called
parallel tempering algorithm, also called random exchange method or multiple
Markov-chain method~\cite{SwWa86,raex}.

In a parallel tempering simulation, $N$ replicas of the statistical system are
simulated in parallel at different temperatures. From the low temperature which
is the target of the investigation, a tower of intermediate temperatures is
built up to a point, where the configurations can move easily from one valley
of the free energy to an other. Let us denote the inverse of these temperatures
by $\beta_1 < \beta_2 < ... < \beta_N$.  

The parallel tempering method involves two components. One is the update of
each individual replica, independently of the others, using a standard
algorithm, e.g. local Metropolis updates. The other are updates that allow to
swap two configurations between  neighboring temperatures.  With this step,
configurations can travel from low to high temperatures and back and thereby
bypass the barriers at low temperatures.

In most implementations, this is realized by proposing the exchange of the field
configurations at inverse temperatures  $\beta_i$ and $\beta_{i+1}$. This is
accepted with the probability 
\begin{equation}
 A_\mathrm{swap} = \mbox{min}[1,\exp((\beta_{i+1} - \beta_{i}) (H_{i+1}  - H_i))] \ ,
\end{equation}
where $H_{i+1}$ and $H_i$ are the values of the Hamiltonian of the replica at
$\beta_{i+1}$ and $\beta_{i}$, respectively. For most systems, this is an
inexpensive step, however, if the temperatures are chosen too far apart, the
acceptance rate will very quickly drop to zero. Therefore, the $\beta_i$  have
to be chosen close enough  such that the acceptance rate $\langle
A_\mathrm{swap} \rangle \gtrsim 0.1$.  In particular for large systems,
this can require the use of large numbers of intermediate temperatures, 
at each of which a replica has to be simulated.

Here we discuss modifications of the replica exchange step of the algorithm.
The basic idea is to not simply swap the two replica, but to 
modify them in such a way as to improve the probability that the move
is accepted.
These modifications  lead  to a higher acceptance rate at given 
differences $\beta_{i+1}-\beta_i$ or the same acceptance rates can be 
achieved using larger differences in the inverse temperature. 

We like to mention that in \cite{OpSc01,BaCh09} annealed swapping is used 
to this end. The authors of \cite{OpSc01}, who have simulated the 
two-dimensional XY model on the square lattice, could indeed reach larger steps
in the temperature, however, this progress does not compensate for 
the additional effort needed for the auxiliary update steps. The authors 
of \cite{BaCh09} have performed a molecular dynamics simulation of 
10 classical non-interacting particles in a potential with several 
local minima. They find that annealed swapping ``is able to achieve the 
computational efficiency of ordinary replica exchange, using fewer replicas.''

In this work, we shall discuss these methods for the example of the
three-dimensional Heisenberg spin glass, however, they are quite general
and can be easily adapted to other models. The Monte Carlo simulation of spin
glasses in general is quite challenging. For the three-dimensional Ising
spin glass at the transition temperature and at temperatures below,  
only lattices up to $32^3$ have been simulated \cite{Janus}.
While there is consensus that the model undergoes a second order phase transition, the error bars of critical
exponents are quite large.

In this letter, we discuss the case of the  physically more realistic
Heisenberg spin glass, where even the nature of the phase transition is still
under debate. Recent works are \cite{FeMaPeTaYo09,ViKa09}.

We consider the Heisenberg spin glass on a simple cubic lattice with
periodic boundary conditions. The classical Hamiltonian is given by
\begin{equation}
 H = - \sum_{<xy>} J_{xy} \; \vec{s}_x \cdot \vec{s}_y \;,
\end{equation}
where the field variables $\vec{s}_x$ are unit vectors with three real 
components; $x$ and $y$
denote the sites of the  lattice. The summation runs over pairs $<\! xy\! >$ of
nearest neighbor sites. 
The $J_{xy}$ are nearest neighbor
interactions with a Gaussian distribution of zero mean and standard
deviation unity. For each set of these quenched interactions,
the expectation values of the observables are computed and then
averaged over different realizations of the $J_{xy}$.

\section{Observables}

In order to study the performance of the algorithm, we have measured
the overlap susceptibility, which is constructed from the overlap variable
\begin{equation}
 q_{\alpha,\beta,x} = s^{(1)}_{\alpha,x} s^{(2)}_{\beta,x}  \ ,
 \label{eq:over}
\end{equation}
where $\vec{s}_x^{\, (1)}$ and $\vec{s}_x^{\, (2)}$ are  the fields at the 
site $x$ of two statistically independent configurations $\{\vec{s}\,\}^{1}$ 
and $\{\vec{s}\,\}^{2}$.
The overlap susceptibility is then given by
\begin{equation}
 \chi  = \frac{1}{L^3} \sum_{\alpha,\beta} 
 \left[ \sum_x q_{\alpha,\beta,x} \right]^2 \ .
\end{equation}
Furthermore, we have measured the internal energy defined by
\begin{equation}
 E = \frac{1}{L^3} \sum_{<xy>} J_{xy} \; \vec{s}_{x} \cdot \vec{s}_{y} \;\;.
\end{equation}

In a physics study of the model, one would consider a larger list of 
quantities, including e.g. the second moment correlation length and various 
cumulants. Also one might study so called chiral quantities; 
see Refs. \cite{FeMaPeTaYo09,ViKa09} for their definition.

To get the two independent configurations required by Eq.~(\ref{eq:over}),
we simulated, as it is usually done, two copies of the system 
for each of the temperatures.
Alternatively one might simulate a single copy and store the configurations
on disk. Then one can combine configurations that are separated 
by $t \gg \tau$ in the Markov chain, where $\tau$ is the autocorrelation 
time.

\section{Improved Replica Exchange}

In the following two sections, we describe two methods, which improve on the
traditional replica exchange between neighboring temperatures. The first is a
decimation procedure, where we study the update under an effective action in
which half the fields have been integrated out. It is described 
in Sec.~\ref{sec:dec}. In the second procedure, given in Sec.~\ref{sec:smear},
we apply an invertible cooling/heating  transformation of the fields,
which leads also to a higher acceptance in the exchange step.
For completeness, we give the details of the hybrid-overrelaxation algorithm
used to simulated the individual replica between the exchange steps
in Sec.~\ref{sec:hor}.

\subsection{Decimation\label{sec:dec}}
The method described in this section is applicable to general models, where
a sub-set of lattice points can be chosen such, that they do not interact
among each other.
Since the generalization to other models, e.g. the Ising spin glass,
is trivial, 
we stay in our discussion with the Heisenberg spin glass model on the 
simple cubic lattice. The partition function in terms of the action 
\begin{equation}
S[\beta,\{\vec{s}\,\}]=\beta H[\{\vec{s}\,\}] = - \beta \sum_{<xy>} J_{xy} \;  
\vec{s}_x \cdot \vec{s}_y 
\end{equation}
is given by
\begin{equation}
Z = \prod_x \left[\int \mbox{d} \vec{s}_x\right] \exp(-S[\beta,\{\vec{s}\,\}]) \;.
\end{equation}
The simple cubic lattice can be divided into two 
sub-sets of points, one called white ($W$), the other black ($B$), such that 
the black points have only white neighbors and vice versa. 
This allows us to write the partition function as 
\begin{equation}
 Z = \prod_{x \in W}  \left[\int \mbox{d} \vec{s}_x \right] 
     \prod_{x \in B} \left[\int \mbox{d} \vec{s}_x  \right]  
      \exp(-S[\beta,\{\vec{s}\,\}]) \ ,
\end{equation}
where now all the integrations over the fields on the black sites 
can be performed:
\begin{equation}
\label{xxxY}
  \prod_{x \in B} \left[\int \mbox{d} \vec{s}_x \right]  
  \exp(-S[\beta,\{\vec{s}\,\}]) 
 = \prod_{x \in B} \left[ \int \mbox{d} \vec{s}_x 
  \exp(\beta \; \vec{s}_x \cdot \vec{S}_x)
\right]  
 =  \prod_{x \in B} I(\beta \; \vec{S}_x) 
\end{equation}
where 
\begin{equation}
\label{bigS}
\vec{S}_x = \sum_{y.nn.x} J_{xy} \; \vec{s}_y
\end{equation}
 is the sum over the fields on
the nearest neighbor sites of $x$. In Eq.~(\ref{xxxY}) we have introduced the 
abbreviation 
\begin{equation}
I(\beta \; \vec{S}_x) =\int \mbox{d} \vec{s}_x \exp(\beta \; 
\vec{s}_x \cdot \vec{S}_x) \; .
\end{equation}
In the case of the Heisenberg model, this  integral can be easily 
performed
\begin{eqnarray} 
\label{HI}
\int \mbox{d} s_{1,x} \int \mbox{d} s_{2,x}  \int  \mbox{d} s_{3,x} 
 \; \delta(\vec{s}_x^{\;2} -1) \; \exp(R_x s_{1,x}) &=&  
c \int_{-1}^{1}  \mbox{d} s_{1,x} \; \exp(R_x s_{1,x}) \nonumber \\
=  \frac{2 c}{R_x} \sinh(R_x) \ ,
\end{eqnarray}
where we have rotated the problem such that $\vec{S}_x$
is a multiple of $(1,0,0)$ and $R_x=\beta |\vec{S}_x|$. 
Putting everything together, the partition function reads
\begin{equation}
 Z =  \prod_{x \in W}  \left[\int \mbox{d} \vec{s}_x \right] 
    \exp(-\tilde S[\beta, \{\vec{s}\,\}_W] )
\ \ \text{with} \ \
\tilde S[\beta, \{\vec{s}\,\}_W] = - \sum_{x \in B} \ln I(\beta \; \vec{S}_x) \; ,
\end{equation}
where the subscript $W$ indicates that $\tilde S$ depends only on the fields
on the white sites.

Now we perform a tempering step with the fields on the white sites only, 
using the action $ \tilde S(\beta, \{\vec{s}\,\}_W) $. 
Given a field $\{\vec{s}\,\}_W^1$ at $\beta_1$ and 
a field $\{\vec{s}\,\}_W^2$ at $\beta_2$, the proposal is to swap 
to  $\{\vec{s}\,\}_W^2$ at $\beta_1$ and $\{\vec{s}\,\}_W^1$ at $\beta_2$.
The acceptance probability for this swap is given by
\begin{equation}
 A_\mathrm{swap} =
\mbox{min} \left[1, \frac{\prod_{x \in B} I(\beta_2 \vec{S}_x^{(1)})}
                         {\prod_{x \in B} I(\beta_1 \vec{S}_x^{(1)})}
                    \frac{\prod_{x \in B} I(\beta_1 \vec{S}_x^{(2)})}
                         {\prod_{x \in B} I(\beta_2 \vec{S}_x^{(2)})}
  \right ] \;.
  \label{eq:aswap}
\end{equation}

In the case of the Heisenberg model, the evaluation of this expression 
is relatively simple. First we notice that the prefactors $\frac{2 c}{R_x}$
cancel. Therefore it remains to evaluate $\prod_{x \in B} \sinh(R_x)$,
for which details are given in Appendix~\ref{app}.

In principle, one could also perform the updates using the decimated 
action  $\tilde S(\beta, \{\vec{s}\,\}_W)$. However there is no efficient 
update for this action. The best idea might be to perform local 
Metropolis updates of $\{\vec{s}\,\}_W$. This would require to evaluate 
$\Delta \tilde S(\beta, \{\vec{s}\,\}_W)$, which is relatively expensive.

Instead, after the tempering step, we insert the fields 
on the black sites again, using their (local) Boltzmann weight. Technically, 
this is done in exactly the same way as a heatbath update is performed.
Having restored the fields on the black sites, we can perform overrelaxation 
and heatbath sweeps as usual.

In our simulations, in order to save CPU-time, we only insert new 
fields on the black sites
if the swap is accepted, otherwise the fields keep their old values. 
Furthermore, we alternate the role of black and white sites from one 
pair of $\beta$-values to the next. In the case of the first pair, we 
chose randomly whether the fields on black or white sites are decimated.

\subsection{Cooling and Heating\label{sec:smear}}

Let us now turn to the  second idea to improve the replica exchange step. 
Inspired by the field transformations proposed in the framework of the Hybrid Monte Carlo 
algorithm \cite{Luscher:2009eq}, it improves on the standard
step, which exchanges just the configuration $(\{\vec{s}\,\}^{1},\{\vec{s}\,\}^{2}) 
\to (\{\vec{s}\,\}^{2},\{\vec{s}\,\}^{1})$ evaluating the action
at the respective other parameters by applying an invertible field transformation
to the configurations
\[
(\{\vec{s}\,\}^{ 1},\{\vec{s}\,\}^{ 2}) \to
 ({\cal F} (\{\vec{s}\,\}^{2}),{\cal F}^{-1} (\{\vec{s}\,\}^{1})) \ .
\]
This can be successful, if we manage to find a transformation, which transforms
a ``typical'' configuration from temperature No. 1 into one more like
those at temperature No. 2 and vice versa.
Obviously this update is reversible. For the acceptance probability
one has to take the Jacobian determinant $\det J_{\cal F}(\cdot)$ 
of the transformation into account. For 
a general transformation we can then use the 
acceptance probability
\begin{equation}
 A_\mathrm{swap} = \mbox{min}\left [1,\frac{\det J_{\cal F}(\{\vec{s}\,\}^{ 2})
 \exp\left \{-\beta_1 H[{\cal F} (\{\vec{s}\,\}^{ 2})]
 -\beta_2  H[{\cal F}^{-1} (\{\vec{s}\,\}^{ 1})] \right \} }
 { \det J_{\cal F}(\{\vec{s}\,\}^{ 1})  \exp\left \{-\beta_1 H[\{\vec{s}\,\}^{ 1}]
  -\beta_2  H[ (\{\vec{s}\,\}^{ 2})]  \right \} } \right] \;.
\end{equation}

For most transformations, computing the Jacobian is a very cumbersome task, it is therefore
advisable to use a transformation, which is composed of elementary steps 
$f$, which only manipulate
one field variable at a time and only depend on its nearest neighbors. 
Then the Jacobian matrix $\partial s'/\partial s$, where $s'=f(s)$, 
can be easily computed along with its determinant.

For the Heisenberg spin glass, 
we propose a transformation which is cooling the configuration when
moving towards a lower temperature and heating it up when moving towards
a higher one.
The specific $f$ we tested here is given by
\[
\vec{s}^{\, \prime}_x = \vec{s}_x \cos \alpha + \frac{\vec{p}}{|\vec{p}\, |} \, \sin \alpha 
\]
with $\alpha=\epsilon |\vec{p}|$ where $\epsilon$ is a tunable parameter.
$\vec{p}$ is the vector in the $\vec{s}_x$--$\vec{S}_x$ plane orthogonal to $\vec{s}_x$
\[
\vec{p}=\vec{S}_x-(\vec{S}_x\cdot \vec{s}_x\, ) \, \vec{s}_x
\]
with $\vec{S}_x$ as defined in Eq.~(\ref{bigS}).
In case of cooling, it reduces the angle between the $\vec{S_x}$ and $\vec{s}_x$ by $\alpha$.
For the inverse operation, a non-linear equation has to be solved. This can be
achieved by a simple Newton iteration which converges very quickly.

In order to compute the Jacobian of the transformation $f$,
we rotate the coordinate system of the integration over $\vec{s}_x$
such that the $z$-axis is parallel to $\vec{S}_x$.
Since only the angle $\theta$ between $\vec{s}_x$ and $\vec{S}_x$
is altered, the relevant integration measure is $\mathrm{d} \cos \theta$. 
For the angle after the cooling $\theta'=\theta-\alpha$, we therefore have
\[
\mathrm{d} \cos \theta'=
\mathrm{d} \cos\theta \left|1-\frac{\mathrm{d}\alpha}{\mathrm{d}\theta}
\right|  (\cos\alpha-\cot\theta \sin\alpha  )
=\mathrm{d} \cos\theta |1-\epsilon \vec{S}_x\cdot \vec{s_x}|
(\cos\alpha-\cot\theta \sin\alpha  )
\]
and we get for the Jacobian determinant
\[
\det J_f(\vec{s}_x)=|1-\epsilon \vec{S}_x\cdot \vec{s_x}|
(\cos\alpha-\cot\theta \sin\alpha  ) \;,
\] 
which has to be accumulated multiplicatively over all steps of the 
cooling/heating in order to get the aggregated value for the whole sweep.

One might expect that the optimal value of the parameter
$\epsilon$ depends on the pair of temperatures.
However, to keep things simple, we have used the same value of $\epsilon$
for all of them. Since  $\beta_{i+1}/\beta_i$ decreases with 
increasing lattice size $L$, also the optimal value of $\epsilon$ is decreasing 
with increasing $L$.  In order to tune $\epsilon$ we have monitored the 
acceptance rate $A$. Reasonable estimates of $A$ can already be obtained from
rather short runs; Here we performed runs with 10000 cycles each.
The optimal value does not depend strongly on the particular set of 
coupling constants.

\subsection{Heat-bath and overrelaxation updates\label{sec:hor}}
In an elementary step of the algorithm, the field at a single site of 
the lattice is updated. Using these updates, we sweep through the lattice 
in typewriter fashion. To this end we use heat-bath and overrelaxation updates:
In the case of the heat-bath update, we chose the component of the new 
field that is parallel to the nearest neighbor sum $\vec{S}_x$, defined in 
Eq.~(\ref{bigS}), as
\begin{equation}
 s_x^{(p)} =  \ln(z + (z^{-1} - z) r)/|\beta\vec{S}_x| \ ,
\end{equation}
where $z=\exp(-\beta|\vec{S}_x|)$ and $r$ is a random number that is uniformly 
distributed in $[0,1]$. The two orthogonal components
\begin{equation}
 s_x^{(o,1)} = \sqrt{1-(s_x^{(p)})^2} \sin \phi  
\;\;,\;\;\; s_x^{(o,2)} = \sqrt{1-(s_x^{(p)})^2} \cos \phi 
\end{equation}
where $\phi$ is uniformly distributed in $[0,2\pi]$.
In an elementary overrelaxation update the field $\vec{s}_x$ is replaced by
\begin{equation}
 \vec{s}^{\, \prime}_x  = 2 \frac{\vec{s}_x \cdot \vec{S}_x}{\vec{S}_x^2} 
 \vec{S}_x  - \vec{s}_x  \;.
\end{equation}
The overrelaxation update takes considerably less CPU-time than the heat-bath 
update, since it requires neither random numbers nor the evaluation
of transcendental functions. The overrelaxation update 
by itself is not ergodic, since it keeps the energy constant. Therefore it 
has to be supplemented by Metropolis, or as it is the case here, heat-bath
updates. It has been demonstrated that such a hybrid of heat-bath and 
overrelaxation updates is clearly more efficient than heat-bath updates 
alone. For a discussion see for example section IV of \cite{FeMaPeTaYo09}.

\section{Numerical results}
In order to test the performance of the algorithm, we have performed 
simulations for $L=16$, $24$ and $32$. Setting up the simulation,
we closely follow Ref. \cite{FeMaPeTaYo09}:  In the tempering 
algorithm, we simulate temperatures $T=1/\beta$ from $T_\mathrm{min}=0.12$ 
up to $T_\mathrm{max}=0.19$. The intermediate temperatures are given by
\begin{equation}
T_i = T_\mathrm{max} \left( \frac{T_\mathrm{min}}{T_\mathrm{max}} \right)^{(i-1)/(N_T-1)}
\end{equation}
where $i=1,2,...,N_T$, with
$N_T=15,27$ and $43$ is used  for $L=16$, $24$ and $32$, 
respectively.  
For each tempering update, 
we perform a heat-bath sweep followed by $\frac{5}{4} L$
overrelaxation sweeps, again as in Ref.~\cite{FeMaPeTaYo09}.
We have not used improvements of this strategy studied in 
Refs.~\cite{KaTrHuTr06,BiNuJa08,HaDiKa10}, since they are complementary to the 
ones discussed here. 
In our implementation, a heat-bath sweep 
takes about 5 times more CPU-time than an overrelaxation sweep. In the 
case of the standard tempering update the CPU-time needed is small compared
with that needed for a heat-bath sweep. For the decimation, 
the tempering update takes a little less CPU-time than a heat-bath sweep, while
for the cooling/heating method the tempering update takes about 
twice the CPU-time of a heat-bath sweep. In particular in the case of the 
cooling/heating method, we did not spent much time on optimizing our 
implementation. In both cases, the CPU-time taken by the tempering update is 
still clearly smaller than that required by the total of the heat-bath and the
overrelaxation sweeps.

In the following, we will use the acceptance rate of the replica exchange
step, the round trip time of the replicas and the auto-correlation time
of the overlap susceptibility as figures of merit for the performance
of the algorithm. Since they might
depend strongly on the particular set of couplings  $\{J_{ij}\}$,
it is quite important to test all variants of the parallel tempering 
algorithm on the same sets.
For each of the lattices sizes, we have therefore generated ten 
realizations of the $\{J_{ij}\}$, on which we performed our tests.

In the case of the standard and the decimation tempering, we have performed
100000, 200000 and 500000 update cycles for each  $\{J_{ij}\}$ for
$L=16$, $24$ and $32$, respectively. For the cooling/heating method
we have performed 500000 update cycles throughout. These numbers are 
clearly larger than the number of update cycles required for equilibration.
In the following, we use $\epsilon=0.017$ for $L=16$ and $\epsilon=0.01$
 for $L=24$.

\subsection{Acceptance rates of the tempering method}
In table \ref{acceptance}, we have summarized our results for the 
acceptance rates of the tempering update. We find that for the given 
choice of $\beta_i$,  the acceptance rates can approximately be 
doubled by our improved  tempering methods.  In almost all cases, the 
acceptance rate slightly increases with increasing $\beta$; only in the 
case of the cooling/heating  method there is a decrease  for $L=16$.
It seems that our choice of the parameter  $\epsilon$ of the cooling/heating
procedure for $L=16$ is better suited for small values of $\beta$
than for large ones. For $L=24$, the situation is just the opposite.
It seems to be beneficial, 
to tune $\epsilon$ for each pair of $\beta$ values separately.

In the case of $L=24$, we tried to determine by how much we can reduce $N_T$
in the case of the improved tempering, here only considering  decimation.
For two of the coupling sets, 
we performed runs using $N_T=22$ instead of $N_T=27$.  
We find acceptance rates of $A \approx 0.146$ up to $A \approx 0.163$, i.e. 
still a bit larger than with the standard tempering and $N_T=27$.

\begin{table}
\begin{ruledtabular}
\begin{tabular}{cccc}
    $L$     &  standard &  decimation &   cool/heat  \\
\colrule
   16  &0.114 --- 0.122  & 0.237 ---  0.256 & 0.290 --- 0.249 \\
   24  &0.116 --- 0.126  & 0.239  --- 0.262 & 0.240 --- 0.320 \\
   32  &0.134 --- 0.145  & 0.262 --- 0.284 &   \\
\end{tabular}
\end{ruledtabular}
\caption{ \label{acceptance}
We give the acceptance rates for the standard and our two improved 
tempering methods as a function of the lattice size. The acceptance rates
depend mildly on the pair of $\beta$-values of the swap. In all cases it
is a monotonic function of $\beta$.  In the table we give the acceptance rate
for the pairs with the smallest and largest values of $\beta$. 
}
\end{table}

\subsection{Round trip and autocorrelation times}
Our next task is to determine, whether these increased acceptance rates 
lead to smaller autocorrelation times. To this end, we have studied 
the round trip time and the integrated auto correlation times  
of the overlap susceptibility at the lowest temperature 
$T_\mathrm{min}=0.12$.

The round trip time is defined in the following way:  we count how often
a configuration runs from the highest temperature $T_\mathrm{max}$ to the 
lowest temperature $T_\mathrm{min}$  and back to $T_\mathrm{max}$.  
This is done for
all  $2 N_T$ configurations. The round trip time $t_R$ is then given as 
the number of all sweeps divided by the number of round trips.

In the case of $L=16$, the round-trip times for the standard
algorithm range from $t_R=2495(46)$ up to $2942(53)$.  These times
are reduced to $t_R=1147(12)$ up to
$1467(19)$ by the decimation method. We have computed the speed-up factor
$t_{\rm R,standard}/t_{\rm R,decimation}$ for each of the ten coupling sets.
The average  is $2.1$. The round-trip times for
the cooling/heating procedure give $t_R=901(4)$ to  $t_R=1170(6)$,
which gives comparable speed-ups of 2.6.

For $L=24$, the round-trip times for the standard algorithm range from
$t_R=9332(209)$ up to $15485(593)$. For the decimation, we get
$t_R=4369(87)$ up to $8100(165)$.
The average speed-up is $1.9$. The cooling/heating
gives round-trip times of $t_R=4038(52)$ to $7437(152)$ and a speed-up of $2.0$.

For $L=32$, the round-trip times for the standard algorithm range from
$t_R=27451(374)$ up to $46206(1452)$. For the improved tempering we get
$t_R=16356(131)$  up to $27432(843)$. With an average speed-up is $1.7$.

The statistical errors that we quote above are only rough estimates, since
they are obtained by blocking the whole data set in ten sub-ensembles.
The fluctuations between different coupling sets are larger than these
individual errors. Fortunately, the relative speed-ups fluctuate
only mildly and can therefore be determined to about $10\%$ accuracy.

\begin{table}
\begin{ruledtabular}
\begin{tabular}{rllllll}
		     & \multicolumn{2}{c}{acc. rate}
		     &  \multicolumn{2}{c}{round trip}
                     & \multicolumn{2}{c}{$\tau_\mathrm{int}(S)$} \\
 \phantom{00}$L$     &  decim. & cool/heat &  decim. & cool/heat &  decim. & cool/heat  \\ 
\colrule
   16                &  2.09   &  2.42 &  2.1   & 2.6 &2.3    & 2.5  \\
   24                &  2.07   &  2.33 &  1.9   & 2.0 &  2.3    & 2.5 \\
   32                &  1.96  &  ---   &  1.7   & --- &  2.4    & --- \\
\end{tabular}
\end{ruledtabular}

\caption{ \label{speedchi}
 Speed-up for the decimation and the cooling/heating procedure
with respect to the standard parallel tempering, given in 
terms of the acceptance rate of the exchange step,  the round-trip time
and the auto-correlation time of the overlap susceptibility.
The error on these numbers is around 2 on the last digit.
}
\end{table}

Next we have computed integrated autocorrelation times defined by
\begin{equation}
 \tau_{\mathrm{int}} = \frac{1}{2} + \sum_{t=1}^{t_f} \rho(t) \;,
\end{equation}
where $\rho(t)$ is the normalized autocorrelation function defined by
\begin{equation}
 \rho(t) = \frac{\langle O(i) O(i+t)  \rangle - \langle O \rangle^2}
                {\langle O^2  \rangle - \langle O \rangle^2}
\end{equation}
and the upper end of the summation is chosen self-consistently 
as $t_f=c \tau_{\mathrm{int}}$. Since the integrated autocorrelation times 
of the overlap susceptibility are larger than those of the 
internal energy, we shall only discuss the former in the following.
We have computed $\tau_{\mathrm{int}}$ for the three choices 
$c=4$, $6$ and $10$.
The extracted auto-correlation times differ considerably due to rather long 
tails in the auto-correlation functions.
The variation of the integrated autocorrelation time over the 
different coupling sets is similar to that of  the round-trip times.
On the other hand, the relative speed-up in the autocorrelation times 
that is obtained by the improved tempering methods depends  only little
of the parameter $c$ and the coupling set. In table \ref{speedchi} we
give the speed-ups obtained with  $c=4$. The improvement found here is 
somewhat larger than that seen in the round trip times.

\subsection{Equilibration}
The equilibration time is an important quantity in spin glass simulations,
because in order to perform the averages over different coupling sets,
many ensembles have to be simulated, in particular since the variation
of interesting observables turns out to be large.
Unfortunately, we were not able to systematically study the equilibration.
To get an impression, we have focussed on the coupling set 5 for $L=16$.
This coupling set has the largest round-trip time among the 10 sets that 
we have studied. 

\begin{figure}
\begin{center}
\scalebox{0.5}
{
\includegraphics{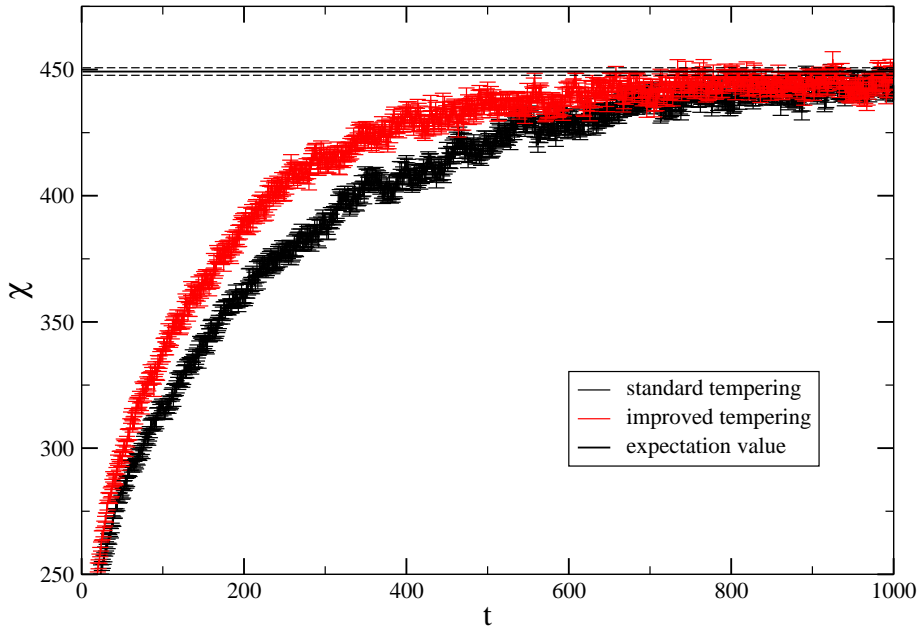}
}
\end{center}
\caption{
\label{EQUIPLOT}  (Color online) We plot the overlap susceptibility  $\chi$
at $T=0.12$ and $L=16$ with the coupling set 5 as a function of the 
Monte Carlo time. We have averaged over 2000 independent runs. 
All runs were started with  $\vec{s}_x  = (1,0,0)$. The straight 
lines give our estimate for the expectation value of $\chi$ and its 
statistical error, obtained from the simulations discussed in the 
previous section. The equilibrium value is approached faster using the 
improved tempering than using the standard one. 
}
\end{figure}

We did the first 1000 iterations of the update 2000 times with different 
random seeds. All these 2000 simulations are started with 
$\vec{s}_{x} = (1,0,0)$.  In figure \ref{EQUIPLOT} we give the averages
of the overlap susceptibility for $T=0.12$ as a function of the 
Monte Carlo time. We see a clear speed-up comparing the standard tempering
simulation with the improved tempering one, e.g. the value 400 is reached 
at $t \approx 230$ in the case of the improved simulation, 
while in the case of the 
standard simulation this is the case for $t \approx 370$. 
These 2000 simulations did cost about 4 days of CPU time.  Therefore we 
abstained from redoing such simulations for all our coupling sets and for  
larger values of $L$. 

\section{Summary and Conclusions}
In this letter, we discuss two methods to improve on the replica exchange
step of parallel tempering. The key idea in both is not to leave the 
configurations as they are and try to swap them between temperatures,
but to transform them during this step. In both methods, we 
find an improvement of the step by roughly a factor of two.

\begin{acknowledgments}
This work was supported by the DFG under the grant No HA 3150/2-1 and
through the SFB/TR 9.
\end{acknowledgments}

\appendix
\section{\label{app}Evaluation of  Eq.~(\ref{eq:aswap})}
Here we give details of the fast numerical evaluation of  $\prod_{x \in B} \sinh(R_x)$,
which appears in Eq.~(\ref{eq:aswap}).
First we write 
\begin{equation}
 \prod_{x \in B} \sinh(R_x) = \exp(\sum_{x \in B} \ln[\sinh(R_x)] )
\end{equation}
for which we need to compute $\ln[\sinh(R_x)]$ efficiently. We use
\begin{equation}
\ln[2 \sinh(R_x)]  = R_x + \ln[1-\exp(-2 R_x)]
\end{equation}
and proceed, depending on the value of $R_x$ in the following way:
For $R_x \gtrapprox 17.5$ , within the numerical precision of 
double precision numbers  $\ln[1-\exp(-2 R_x)] =0$.  For $1 < R_x < 17.5$, 
we have used a pre-computed table, for $R_x=0.9,1.0,...
,17.5,17.6$  In order to get 
$\ln[1-\exp(-2 R_x)]$ for any $R_x$ in $1 < R_x < 17.5$, we have quadratically 
interpolated the entries of this table.  We have checked that the error
of this evaluation is at most of the order $10^{-16}$.
If $R_x<1$, which very rarely happens in the simulation, we have used the 
functions of the C-library to compute  $\ln[1-\exp(-2 R_x)]$.


\begin{thebibliography}{99}

\bibitem{raex} C. J. Geyer in {\em Computer Science and Statistics: Proc. of
 the 23rd Symposium on the Interface}, edited by E. M. Keramidas \/
(Interface Foundation, Fairfax Station, 1991), p. 156; K. Hukushima and
 K. Nemoto, J.\ Phys.\ Soc.\ Jpn.\ {\bf 65}, 1604 (1996);
for a review, see
D. J. Earl and M. W. Deem, Phys.\ Chem.\ Chem.\ Phys.\ {\bf 7}, 3910 (2005).

\bibitem{SwWa86}
R. H. Swendsen, J. S. Wang, 
{\sl Replica Monte Carlo Simulation of Spin-Glasses},
Phys.\ Rev.\ Lett.\  {\bf 57}, 2607 (1986).

\bibitem{OpSc01}
S. B. Opps and J. Schofield,
{\sl Extended state space Monte-Carlo methods}, 
Phys.\ Rev.\ E {\bf 63}, 056701 (2001).

\bibitem{BaCh09}
A. J. Ballard and C. Jarzynski,
{\sl Replica exchange with nonequilibrium switches},
P.\ Natl.\ Acad.\ Sci.\ USA {\bf 106}, 12224 (2009).


\bibitem{Janus}
R. Alvarez Banos et al. [Janus Collaboration], 
{\sl Reliable determination of the order parameter for the $D=3$ Ising 
spin glass}, [arXiv:1003.2943].

\bibitem{FeMaPeTaYo09}
L. A. Fernandez, V. Martin-Mayor, S. Perez-Gaviro, A. Tarancon, and A. P. Young,
{\sl Phase transition in the three dimensional Heisenberg spin glass:
 Finite-size scaling analysis}, [arXiv:0905.0322v2],
 Phys.\ Rev.\ B {\bf 80}, 024422 (2009).

\bibitem{ViKa09}
D. X. Viet, H. Kawamura,
{\sl Monte Carlo studies of the chiral and spin orderings of the
three-dimensional Heisenberg spin glass}, [arXiv:0904.3699],
Phys.\ Rev.\ B {\bf 80}, 064418 (2009).

\bibitem{Luscher:2009eq}
M. L\"uscher,
{\sl Trivializing maps, the Wilson flow and the HMC algorithm},
[arXiv:0907.5491],
Commun.\ Math.\ Phys.\  {\bf 293 } (2010)  899.

\bibitem{KaTrHuTr06}  
H. G. Katzgraber, S. Trebst, D. A. Huse, and M. Troyer,
{\sl Feedback-optimized parallel tempering Monte Carlo}, 
[arXiv:cond-mat/0602085],
J.\ Stat.\ Mech.: Theory Exp. {\bf 2006}, P03018.

\bibitem{BiNuJa08}
E. Bittner, A. Nussbaumer, and W. Janke, 
{\sl Make life simple: unleash the full power of the parallel tempering 
algorithm},
[arXiv:0809.0571], 
Phys.\ Rev.\ Lett.\ {\bf 101}, 130603 (2008).

\bibitem{HaDiKa10}
F. Hamze, N. Dickson, and K. Karimi,
{\sl Robust Parameter Selection for Parallel Tempering}, 
[arxiv:1004.2840], 
Int.\ J.\  Mod.\ Phys.\ C {\bf 21} 603 (2010).

\end{thebibliography}
\end{document}